\documentclass[aps,prd,twocolumn,superscriptaddress,nofootinbib,groupedaddress]{revtex4-2}
 
\pdfoutput=1

\usepackage{slashed}
\usepackage{colortbl}
\usepackage{latexsym}
\usepackage{amssymb}
\usepackage{amsmath}
\usepackage{mathrsfs}
\usepackage{graphicx}
\usepackage{xcolor}
\usepackage{adjustbox}
\usepackage{easybmat}
\usepackage{booktabs}
\usepackage{multirow}
\usepackage{rotating}
\usepackage{soul}
\usepackage{hyperref}
\usepackage{mathtools}
\usepackage{calrsfs}

\newcommand{\ii}{\ensuremath{\mathrm{i}}}

\DeclareMathOperator*{\sumint}{
\mathchoice
 {\ooalign{$\displaystyle\sum$\cr\hidewidth$\displaystyle\int$\hidewidth\cr}}
 {\ooalign{\raisebox{.14\height}{\scalebox{.7}{$\textstyle\sum$}}\cr\hidewidth$\textstyle\int$\hidewidth\cr}}
 {\ooalign{\raisebox{.2\height}{\scalebox{.6}{$\scriptstyle\sum$}}\cr$\scriptstyle\int$\cr}}
 {\ooalign{\raisebox{.2\height}{\scalebox{.6}{$\scriptstyle\sum$}}\cr$\scriptstyle\int$\cr}}
}

\definecolor{greenish}{HTML}{458A51}
\definecolor{purplish}{HTML}{95529E}

\begin{document}


\title{New insights into two-loop running in effective field theories}

\author{Mikael Chala}
\email{mikael.chala@ugr.es}
\author{Javier L\'opez Miras}
\email{jlmiras@ugr.es}
\affiliation{Departamento de F\'isica Te\'orica y del Cosmos, Universidad de Granada, E--18071 Granada, Spain}

\begin{abstract} 
\vspace{0.5cm}
We show that, by viewing a 4D effective-field theory as the infrared (IR) limit of the compactified version in 5D, we can compute two-loop anomalous dimensions without gauge-breaking counter-terms, IR re-arrangement or geometric methods. The ultraviolet (UV) divergences in 4D are read from the IR ones in the matching from 5D to 4D.
We use this approach to cross-check recent results in the literature, as well as to compute novel two-loop anomalous dimensions in the SMEFT to dimension eight and certain critical exponents in the charged fixed point of the Abelian Higgs model at large number of flavors.
\vspace{1cm}
\end{abstract}

\maketitle

\newpage

\section{Introduction} 
There is growing interest, both in particle physics and in the study of critical phenomena, in extending the computation of renormalization group equations (RGEs) in effective field theory (EFT) beyond the one-loop level. 
In the first case, they provide a necessary ingredient to confront theoretical predictions with the precise experimental data. This is particularly relevant in studies of the Standard model (SM) EFT~\cite{Brivio:2017vri,Aebischer:2025qhh,Isidori:2023pyp}, where two-loop RGEs have been only recently computed~\cite{Born:2024mgz}. (See also Refs.~\cite{Panico:2018hal,DiNoi:2024ajj,Jenkins:2023bls,Naterop:2024cfx,Naterop:2025cwg,Naterop:2025lzc,Duhr:2025zqw,Banik:2025wpi,DiNoi:2025arz,Zhang:2025ywe,DiNoi:2025tka} for partial results.) They have been also instrumental for unraveling hidden structures in EFT using amplitudes methods~\cite{Bern:2019wie,EliasMiro:2020tdv,EliasMiro:2021jgu}, including non-renormalization theorems up to four loops.

On the side of critical phenomena, high-loop calculations are necessary to accurately determine scaling dimensions and the spectrum of operators at fixed points accessible via $\epsilon$-expansion~\cite{Wilson:1971dc,Wilson:1973jj}. Current computations in scalar field theory reach up to five loops~\cite{FCZhang:1982,Kleinert:1991rg,Cao:2021cdt,Henriksson:2025vyi,Gracey:2025aoj,Henriksson:2025hwi}, and there is strong activity towards reproducing these data on the basis of conformal-field theory (CFT) methods alone~\cite{Rychkov:2015naa,Basu:2015gpa,Roumpedakis:2016qcg,Dey:2017oim,Henriksson:2018myn}.

Yet, two-loop calculations in EFT are notoriously difficult in EFT, particularly in the presence of gauge interactions. This explains in part the long gap between the one-loop determination of the SM EFT RGEs~\cite{Jenkins:2013wua,Jenkins:2013zja,Alonso:2013hga} and the recent two-loop result; as well as the little CFT data in gauge theories, as for example at the charged fixed point in the Abelian Higgs at large number of flavours~\cite{Ihrig:2019kfv}.
In this paper, we present a novel approach to extract two-loop anomalous dimensions in bosonic EFTs. It is based on viewing a 4D EFT as the IR limit of the 5D counterpart upon compactification of the 5th dimension. The IR poles arising from integrating out the Kaluza-Klein (KK) modes match the UV poles in the 4D theory.

The paper is organized as follows. We motivate our approach in sec.~\ref{sec:motivation}, and explain it in detail in sec.~\ref{sec:idea}. We explore different applications in sec.~\ref{sec:applications}. We conclude in sec.~\ref{sec:conclusions}, whereas Appendix~\ref{app:example} is dedicated to highlight some nuances of the method within a concrete detailed example.

\section{Motivation}
\label{sec:motivation}
\begin{figure}[b]
    \includegraphics[width=0.3\columnwidth]{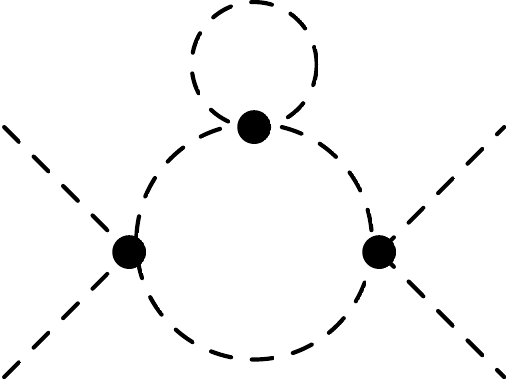}
    \includegraphics[width=0.3\columnwidth]{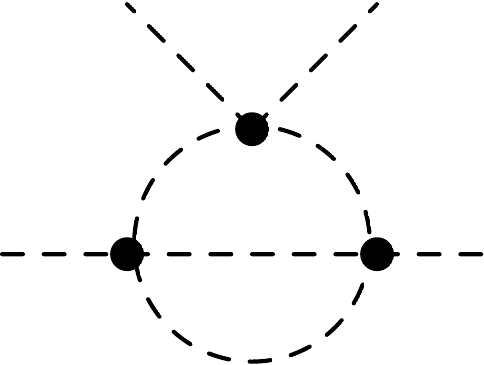}
    {\includegraphics[width=0.3\columnwidth]{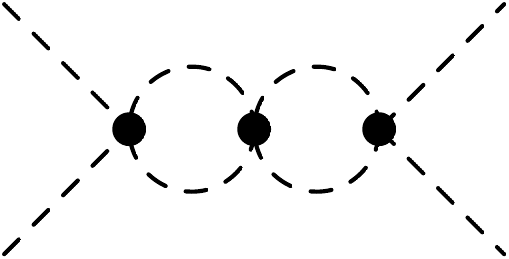}\vspace{0.2cm}}
    \includegraphics[width=0.3\columnwidth]{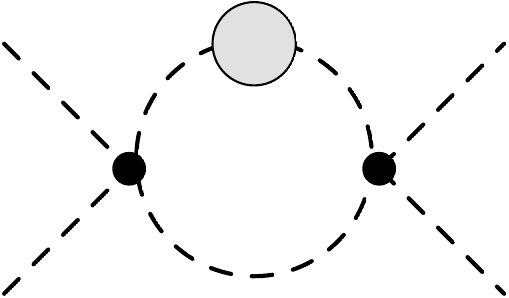}
    \hspace{0.5cm}
    \includegraphics[width=0.3\columnwidth]{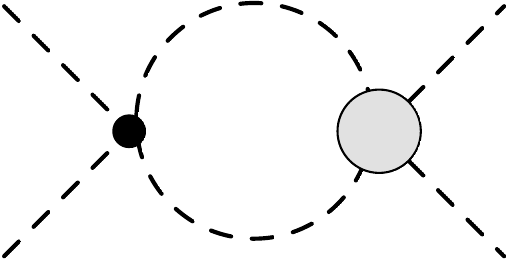}
    \caption{\it Relevant diagrams (up to permutations of external legs) for the two-loop renormalization of the quartic coupling in scalar field theory. Gray blobs represent one-loop counterterms.}\label{fig:diagrams}
\end{figure}
Let us first consider the two-loop renormalization of the quartic coupling in the simplest renormalizable scalar field theory:
\begin{equation}
    \mathcal{L} = \frac{1}{2}(\partial\varphi)^2-\frac{1}{2}m^2\varphi^2-\frac{1}{4}\lambda \varphi^4\,.
\end{equation}
We work in dimensional regularization (dimReg) with space-time dimension $d=4-2\epsilon$. The relevant diagrams are shown in Fig.~\ref{fig:diagrams}, which give:
\begin{widetext}
\begin{align}
    \mathcal{G}_{\varphi\varphi\varphi\varphi} &=36\int_q\left[\lambda^2 \delta_{m^2}^{(1)}\frac{1}{(q^2-m^2)^3}+\lambda\delta_\lambda^{(1)}\frac{1}{(q^2-m^2)^2}\right]\,\nonumber\\
    &-\lambda^3\int_{qr}\left\lbrace\frac{108}{(q^2-m^2)^3 (r^2-m^2)}+\frac{216}{(q^2-m^2)(r^2-m^2)[(q+r)^2-m^2]^2}+\frac{54}{(q^2-m^2)^2 (r^2-m^2)^2}\right\rbrace\,,
\end{align}
\end{widetext}
where
\begin{equation}
    \delta_\lambda^{(1)} = \frac{9}{16\pi^2\epsilon}\lambda^2\,,\quad \delta_{m^2}^{(1)} = -\frac{3}{16\pi^2\epsilon}\textcolor{blue}{m^2}\lambda
\end{equation}
are the one-loop counterterms for $\lambda$ and $m^2$, respectively.

Plugging now the following $\epsilon$-expanded expressions for the loop integrals~\cite{Davydychev:1992mt} (note that we ignore the mass in certain denominators to lighten the expressions),
\begin{align}
    \int_q\frac{1}{q^4} &= \frac{1}{16\pi^2\epsilon}+\cdots\,,\nonumber\\
    \int_q\frac{1}{(q^2-m^2)^3} &= \frac{1}{32\pi^2 \textcolor{blue}{m^2}}\,,\nonumber\\
    \int_{qr} \frac{1}{q^6 r^2} &= -\frac{1}{512\pi^4\epsilon}\,,\nonumber\\
    \int_{qr}\frac{1}{q^2 r^2(q+r)^4} &= \frac{1}{512\pi^4}\left(\frac{1}{\epsilon}+\frac{1}{\epsilon^2}\right)+\cdots\nonumber\\
    \int_{qr}\frac{1}{q^4r^4} &= \frac{1}{256\pi^4\epsilon^2},
\end{align}
(the ellipses encode logarithmic terms),
we obtain:
\begin{equation}\label{eq:example}
   \delta\lambda^{(2)}= \frac{1}{256 \pi^4}\lambda^3 \left(-\frac{54}{\epsilon}+\frac{81}{\epsilon^2}\right)\,.
\end{equation}
(Note that, to get the beta function in the literature, the wavefunction renormalization has to be computed too.)

Some simple observations follow from this calculation. To start with, in the limit $m^2\to 0$, the UV divergences can not be disentangled from the IR ones, because dimReg regulates both with the same $\epsilon$. (This is more acute with non-vanishing external momenta $p$, because the expansion in $q,r\gg p$ introduces spurious IR divergences.)
Moreover, there are sub-divergences, $\sim\frac{1}{\epsilon}\log{m^2}$, that are only canceled by one-loop diagrams involving one-loop counterterms. Most importantly, they correct the pure two-loop $1/\epsilon$ ensuing from the last three diagrams in Fig.~\ref{fig:diagrams}, as a  result of $m^2$ cancellation in $\delta_{m^2}^{(1)}\int_q (q^2-m^2)^{-3}$; see the \textcolor{blue}{blue terms} above. All in all, the result would be wrong if the mass term, which is only in place to regulate IR divergences, is not itself renormalized and its counterterm included in one-loop diagrams.

While this seems innocuous in a renormalizable scalar theory, it becomes dangerous in EFT. This is so because the number of (physical and redundant) higher-dimensional operators, to be included in one-loop diagrams with counterterms, increases rapidly with energy dimension. Things get even worse in the presence of gauge interactions, because gauge boson masses break gauge invariance, requiring gauge-breaking counter-terms in one-loop diagrams and the number of such terms is enormous if renormalizability is abandoned.~\footnote{Note that, even if IR divergences are regulated somehow else, and even if using the background field method~\cite{Abbott:1981ke}, quantum fluctuations of gauge fields break gauge invariance down to BRST only~, requiring the inclusion of counterterms for gauge-breaking BRST-preserving operators with quantum fields, \textit{e.g.} $\mathcal{O}\sim (A^\mu D_\mu \hat{A})^2$ in $SU(2)$, where $\hat{A}$ stands for the background.}

In light of this, several alternative computational approaches have been proposed in the literature, ranging from exact propagator decompositions~\cite{Chetyrkin:1997fm} to infrared rearrangement techniques~\cite{Vladimirov:1979zm,Chetyrkin:1980pr}, with the $R^*$ method~\cite{Chetyrkin:1984xa} being the most prominent example.

\section{4D as the low-energy limit of 5D}
\label{sec:idea}
Alternatively to previous methods for isolating two-loop divergences in dimReg, we propose to see a given theory in $d=4$ dimensions as the IR limit of its counterpart in $D=1+d$ upon compactifying the fifth dimension. This generates a tower of Kaluza-Klein modes of mass $m\sim \pi \Lambda$, where $\Lambda = 1/R$ and $R$ is the size of the extra dimension. They can be integrated out, in the same spirit as dimensional reduction in thermal field theory~\cite{Kajantie:1995dw}. Diagrammatically, this can be done by matching correlators computed in 5D with vanishing momentum along the fifth dimension with correlators computed in 4D.

Just as in any other matching calculation, the $1/\epsilon$ poles in the matching involve IR and UV divergences. The former are precisely the UV poles of the EFT~\cite{Manohar:2018aog} (i.e., the ones we aim to compute in 4D), while the latter vanish for the following two reasons.
(i) One-loop diagrams in odd space-time dimension are not logarithmically divergent, so they give no $1/\epsilon$ poles in dimReg. (ii) With the exception of masses and trilinear scalar interactions, couplings in 5D have negative mass dimension; as a result, combinations of couplings that renormalize a given operator in 4D do not generate UV divergences for the corresponding operator in 5D, but only for operators of higher dimension from the 4D perspective.
Therefore, the 4D UV divergences are simply the IR divergences arising in the matching from 5D to 4D. 
No mass or any other regulators are needed. Moreover, one-loop diagrams with one-loop counterterms vanish, avoiding the computation of redundant one-loop divergences (including gauge-breaking ones). Furthermore, the hard-region expansion does not introduce spurious IR divergences because these are regulated by the mass of KK modes.

To very briefly exemplify how this works, let us recompute Eq.\eqref{eq:example}. We only need to consider the first three diagrams in Fig.~\ref{fig:diagrams}. We obtain the following 5D correlator $\mathcal{G}_{\varphi\varphi\varphi\varphi}$ at zero external momentum:
\begin{align}
    \mathcal{G}_{\varphi\varphi\varphi\varphi} &= -\lambda^3 \sumint_{QR} \left[\frac{108}{Q^6 R^2}+\frac{216}{Q^2 R^2 (Q+R)^4}+\frac{54}{Q^4 R^4}\right]\,.
\end{align}
The sum-integrals, encoding the contributions of the infinite KK modes to the matching, are defined as
\begin{align}
    \sumint_{QR} &\equiv \Lambda^2\sum_{n=-\infty}^\infty \sum_{m=-\infty}^\infty\int_q \frac{d^d q}{(2\pi)^d}\int_r\frac{d^d r}{(2\pi)^d}
\end{align}
where $Q=(q,Q_5)$ and $R=(r,R_5)$ are 5-dimensional loop momenta, with $Q_5=2\pi n\Lambda$ and $R_5=2\pi m\Lambda$; $n,m$ label the KK modes.

Ref.~\cite{Davydychev:2023jto} shows how these integrals factorize into one-loop sum-integrals, which are trivial to solve:
\begin{align}
    I_\alpha^r &\equiv\sumint_Q \frac{Q_5^r}{Q^{2\alpha}} \nonumber\\
    &= \tilde{\mu}^{2\epsilon} \Lambda \frac{(1+(-1)^r)}{(2\pi\Lambda)^{2\alpha-r-d}}\frac{\Gamma(\alpha-d/2)}{(4\pi)^{d/2}\Gamma(\alpha)}\zeta(2\alpha-r-d)\,,
\end{align}
where $\tilde{\mu}^2=(e^{\gamma_E}\mu^2)/(4\pi)$, $\mu$ is the $\overline{\rm{MS}}$ scale and $\gamma_E$ the Euler-Mascheroni constant.

Using the algorithm in Ref.~\cite{Davydychev:2023jto}, we obtain:
\begin{align}
    \mathcal{G}_{\varphi\varphi\varphi\varphi} &= -\lambda^3 \left[108 I_1^0 I_3^0-\frac{216}{(d-5)(d-2)}(I_2^0)^2+54 (I_2^0)^2\right]\,.
\end{align}
Expanding the integrals $I_\alpha^r$ in $\epsilon$ around $d=4-2\epsilon$, we get:
\begin{align}
    \mathcal{G}_{\varphi\varphi\varphi\varphi} &= \frac{1}{128 \pi^4} \lambda^3 \left(\frac{54}{\epsilon}-\frac{81}{\epsilon^2}\right)\Lambda^2\,,
\end{align}
from where we trivially read $\delta_\lambda^{(2)}$.

It is worth pointing out one subtlety of this method. In the presence of gauge bosons $A_\mu^a$, the low-energy limit of the 5D theory involves also adjoint scalars $A_5^a$ associated with the compactified dimension (just like Debye modes in thermal field theory). They can only alter running equations beyond $\mathcal{O}(g^2)$, with $g$ the gauge coupling. In that case, contributions from $A_5$ must be subtracted.

We refer to Appendix~\ref{app:example} for a  detailed example, where we also show explicitly how the operators that run in 5D are precisely those that we are not interested in 4D.

\section{Applications}
\label{sec:applications}
In what follows, we apply this method to the computation of certain two-loop anomalous dimensions in the SMEFT at dimension six, to compare with Ref.~\cite{Born:2024mgz}, as well as, for the first time, to dimension eight. We discuss this latter result in light of custodial symmetry and positivity bounds.

Likewise, we compute, for the first time, the anomalous dimensions of dimension-six terms in the Abelian Higgs model at the charged fixed point present in the limit of large number of scalar flavors.

\subsection{SMEFT RGEs at dimensions six and eight}
\label{subsec:rges smeft}
\begin{table}[t]
 \begin{tabular}{|p{0.3\columnwidth} m{0.4\columnwidth}|}
 \hline
 \rowcolor{gray!20} \multicolumn{2}{|c|}{$\boldsymbol{\phi^2 D^4}$} \\
 \hline
 \hspace{0.1\columnwidth} \textcolor{gray}{$\mathcal{O}_{D\phi}$} & \textcolor{gray}{$D^2\phi^\dagger D^2\phi$} \\
 \hline
 \rowcolor{gray!20} \multicolumn{2}{|c|}{$\boldsymbol{\phi^4 D^2}$} \\
 \hline
 \hspace{0.1\columnwidth} $\mathcal{O}_{\phi\square}$ & $(\phi^\dagger\phi)\square (\phi^\dagger\phi)$ \\
 \hspace{0.1\columnwidth} $\mathcal{O}_{\phi D}$ & \textcolor{gray}{$(\phi D_\mu\phi)^\dagger (\phi^\dagger D^\mu\phi)$} \\
 \hspace{0.1\columnwidth} \textcolor{gray}{$\mathcal{O}^\prime_{\phi D}$} & \textcolor{gray}{$(\phi^\dagger\phi)(D_\mu\phi^\dagger D^\mu\phi)$} \\
 %
 %
 \hline
 \rowcolor{gray!20} \multicolumn{2}{|c|}{$\boldsymbol{\varphi^6}$} \\
 \hline
 \hspace{0.1\columnwidth} $\mathcal{O}_\phi$ & $(\varphi^\dagger\phi)^3$\\
 \hline
 \rowcolor{gray!20} \multicolumn{2}{|c|}{$\boldsymbol{X^2 D^2}$} \\
 \hline
 \hspace{0.1\columnwidth} \textcolor{gray}{$\mathcal{O}_{2B}$} & \textcolor{gray}{$-\frac{1}{2}\partial_\mu B^{\mu\nu} \partial_\rho B^{\rho\nu}$}\\
 \hline
 \rowcolor{gray!20} \multicolumn{2}{|c|}{$\boldsymbol{\phi^2 X D^2}$} \\
 \hline
 \hspace{0.1\columnwidth} \textcolor{gray}{$\mathcal{O}_{BD\phi}$} & \textcolor{gray}{$D_\nu B^{\mu\nu} (\phi^\dagger \ii \overleftrightarrow{D}_\mu\phi)$}\\
 \hline
 \rowcolor{gray!20} \multicolumn{2}{|c|}{$\boldsymbol{X^2\phi^2}$} \\
 \hline
 \hspace{0.1\columnwidth} $\mathcal{O}_{\phi B}$ & $(\phi^\dagger\phi) B_{\mu\nu} B^{\mu\nu}$\\
 \hline
 \end{tabular}
 \caption{\it Bosonic CP-conserving SMEFT operators at dimension six. Redundant ones are in gray. We follow the conventions of Refs.~\cite{Gherardi:2020det}.}\label{tab:smeftdim6}
\end{table}
For simplicity, we consider the bosonic SMEFT Lagrangian in the limit of vanishing $g_2$, $g_3$ and Higgs mass. It reads:
\begin{align}
    \mathcal{L} = &-\frac{1}{4}B_{\mu\nu}B^{\mu\nu} + (D_\mu\phi)^\dagger (D^\mu\phi) - \lambda |\phi|^4 \nonumber\\
    &+ \frac{1}{M_\Lambda^2}\sum_i c_i^{(6)} \mathcal{O}^{(6)}+\frac{1}{M_\Lambda^4}\sum_i c_i^{(8)} \mathcal{O}^{(8)}+\cdots
\end{align}
\begin{table}[t]
\begin{tabular}{|p{0.3\columnwidth} m{0.5\columnwidth}|}
 \hline
 \rowcolor{gray!20} \multicolumn{2}{|c|}{$\boldsymbol{\phi^2 D^6}$} \\
 \hline
 \hspace{0.1\columnwidth} \textcolor{gray}{$\mathcal{O}_{\phi^2 D^6}^{(1)}$} & \textcolor{gray}{$(D^2\phi^\dagger)D_\mu D_\nu D^\mu D^\nu\phi$} \\
 \hline
 \rowcolor{gray!20} \multicolumn{2}{|c|}{$\boldsymbol{\phi^4 D^4}$} \\
 \hline
 \hspace{0.1\columnwidth} $\mathcal{O}_{\phi^4 D^4}^{(1)}$ & $(D_{\mu} \phi^{\dag} D_{\nu} \phi) (D^{\nu} \phi^{\dag} D^{\mu} \phi)$ \\
 \hspace{0.1\columnwidth} $\mathcal{O}_{\phi^4 D^4}^{(2)}$ & $(D_{\mu} \phi^{\dag} D_{\nu} \phi) (D^{\mu} \phi^{\dag} D^{\nu} \phi)$ \\
 \hspace{0.1\columnwidth} $\mathcal{O}_{\phi^4 D^4}^{(3)}$ & $(D^{\mu} \phi^{\dag} D_{\mu} \phi) (D^{\nu} \phi^{\dag} D_{\nu} \phi)$ \\
 \hspace{0.1\columnwidth} \textcolor{gray}{$\mathcal{O}_{\phi^4 D^4}^{(4)}$} & \textcolor{gray}  {$D_\mu\phi^\dagger D^\mu\phi(\phi^\dagger D^2\phi + \text{h.c.})$} \\
 \hspace{0.1\columnwidth} \textcolor{gray}{$\mathcal{O}_{\phi^4 D^4}^{(6)}$} & \textcolor{gray}{{$(D_\mu\phi^\dagger \phi) (D^2\phi^\dagger D^\mu\phi) + \text{h.c.}$}} \\
 \hspace{0.1\columnwidth} \textcolor{gray}{$\mathcal{O}_{\phi^4 D^4}^{(8)}$} & \textcolor{gray}{$(D^2\phi^\dagger\phi) (D^2\phi^\dagger\phi)+\text{h.c.}$} \\
 \hspace{0.1\columnwidth} \textcolor{gray}{$\mathcal{O}_{\phi^4 D^4}^{(10)}$} & \textcolor{gray}{$(D^2\phi^\dagger D^2\phi) (\phi^\dagger\phi)$} \\
 \hspace{0.1\columnwidth} \textcolor{gray}{$\mathcal{O}_{\phi^4 D^4}^{(11)}$} & \textcolor{gray}{$(\phi^\dagger D^2\phi) (D^2\phi^\dagger\phi)$} \\
 \hspace{0.1\columnwidth} \textcolor{gray}{$\mathcal{O}_{\phi^4 D^4}^{(12)}$} & \textcolor{gray}{$(D_\mu\phi^\dagger \phi)(D^\mu\phi^\dagger D^2\phi) + \text{h.c.}$} \\
 \hline
 \rowcolor{gray!20} \multicolumn{2}{|c|}{$\boldsymbol{\phi^6 D^2}$} \\
 \hline
 \hspace{0.1\columnwidth} $\mathcal{O}_{\phi^6 D^2}^{(1)}$ & $(\phi^\dagger\phi)^2(D^\mu\phi^\dagger D_\mu\phi)$ \\
 \hspace{0.1\columnwidth} $\mathcal{O}_{\phi^6 D^2}^{(2)}$ & $(\phi^\dagger\phi)(\phi^\dagger\sigma^I\phi)(D^\mu\phi^\dagger\sigma^I D_\mu\phi)$ \\
 \hspace{0.1\columnwidth} \textcolor{gray}{$\mathcal{O}_{\phi^6 D^2}^{(3)}$} & \textcolor{gray}{$(\phi^\dagger\phi)^2(\phi^\dagger D^2\phi+\text{h.c.})$} \\
 \hspace{0.1\columnwidth} \textcolor{gray}{$\mathcal{O}_{\phi^6 D^2}^{(4)}$} & \textcolor{gray}{$(\phi^\dagger\phi)^2 D^\mu(\phi^\dagger\ii \overleftrightarrow{D}_\mu\phi)$} \\
 %
 %
 %
 %
 %
 \hline
\end{tabular}
\caption{\it CP-conserving Higgs operators at dimension eight with at most six fields. Redundant ones are in gray. We follow the conventions of Refs.~\cite{Chala:2021cgt,Murphy:2020rsh}.}\label{tab:smeftdim8}
\end{table}

\noindent where the dimension-six operators $\mathcal{O}^{(6)}$ and the dimension-eight ones $\mathcal{O}^{(8)}$ are those in Tabs.~\ref{tab:smeftdim6} and \ref{tab:smeftdim8}, respectively. The ellipses represent higher-dimensional terms.

Using our approach, we obtain the two-loop simple poles of the dimension-six operators shown in Eq.~\eqref{eq:results dim 6 SMEFT}, where $c_i,r_i$ stand for the Wilson coefficients (WCs) of physical and redundant operators, respectively. The expressions comprise the two-loop-order results valid up to $\mathcal{O}(g_1^4)$, after performing kinetic normalization, and they are all factored by the standard $1/(256\pi^4\epsilon)$.
\begin{align}
    \tilde{K}_{\phi} &= 3\lambda^2\,,\nonumber\\[0.2cm]
    \tilde{r}_{D\phi} &= -\frac{1}{24}\left[
    (5g_1^2+12\lambda)c_{\phi D}+(5 g_1^2-48\lambda)c_{\phi\square})\right]\,,\nonumber\\[0.2cm]
    \tilde{c}_{\phi\square} &= \frac{1}{24}\bigg[
    95g_1^2\lambda c_{\phi D}\nonumber\\
    &+(984\lambda^2-103g_1^2\lambda)c_{\phi\square}+288\lambda^2 c_{\phi}\bigg]\,,\nonumber\\[0.2cm]
    \tilde{c}_{\phi D} &= \frac{1}{12}\bigg[
    (504\lambda^2-5g_1^2\lambda)c_{\phi D}
    +292g_1^2 c_{\phi\square}\bigg] \,,\nonumber\\[0.2cm]
    \tilde{r}_{\phi D}' &= -\frac{1}{12}\bigg[
    10 g_1^2\lambda c_{\phi D}+(19g_1^2\lambda-384\lambda^2) c_{\phi \square}
    +288 c_\phi\bigg]\,,\nonumber\\[0.2cm]
    \tilde{c}_{\phi} &= 2\bigg [
    (-18g_1^2\lambda+459\lambda^2)c_\phi 
    \nonumber\\
    &+(2 g_1^2 \lambda^2 + 250 \lambda^3)c_{\phi D}+(32 g_1^2 \lambda^2 - 1032 \lambda^3)c_{\phi\square}\bigg]\,,\nonumber\\[0.2cm]
    %
    %
    \tilde{r}_{BD\phi} &= -\frac{1}{12}\left[
    5 g_1\lambda c_{\phi D}+10 g_1\lambda c_{\phi\square}\right]\,,\nonumber\\[0.2cm]
    \tilde{c}_{\phi B} &= \frac{1}{2}(g_1^2 c_{\phi D}+2 g_1^2 c_{\phi\square} + 36 \lambda^2 c_{\phi B})\,.
    \label{eq:results dim 6 SMEFT}
\end{align}
All other remaining WC's vanish. Notice that we only show contributions up to order $g_1^2\lambda^2$, neglecting $\mathcal{O}(g_1^4)$ terms. (Following standard power-counting rules, we assume $c_{\phi B}\sim\mathcal{O}(g_1^2)$ too.) Computing higher-order corrections in $g_1^2$ requires the subtraction of contributions from the adjoint scalar $B_5$ arising in the compactification from 5D to 4D. We explain this nuance in detail in Appendix \ref{app:example}. 

The redundant operators can be removed using the following field redefinitions~\cite{Chala:2024llp}:
\begin{align}
    \tilde{c}_{\phi\square} &\to \tilde{c}_{\phi\square}-\frac{1}{2}g_1^2 \tilde{r}_{2B}+\frac12 \tilde{r}_{\phi D}^\prime+\frac{1}{2}g_1 \tilde{r}_{BD\phi}\,,\nonumber\\[0.3cm]
    \tilde{c}_{\phi D} &\to \tilde{c}_{\phi D}-\frac12 g_1^2 \tilde{r}_{2B}+2 g_1 \tilde{r}_{BDH}\,,\nonumber\\[0.3cm]
    \tilde{c}_{\phi} &\to \tilde{c}_{\phi}+4\lambda^{2}\tilde{r}_{D\phi}+2\lambda\tilde{r}_{\phi D}^\prime\,.
    \label{eq:field redefinitions smeft 6}
\end{align}
All other physical operators remain invariant upon applying these field redefinitions.

After this, we can trivially obtain the two-loop component of the beta functions, that we write in the standard manner
\begin{equation}
    \beta_{c_i} = \frac{1}{(4\pi)^2}\beta_{c_i}^{(1\ell)}+\frac{1}{(4\pi)^4}\beta_{c_i}^{(2\ell)}+\cdots
\end{equation}
where the superindex indicates the loop order and the ellipses stand for contributions arising at three loops and above.

We obtain:
\begin{align}
    \beta_{c_{\phi\square}}^{(2\ell)} &= \left( 22 g_1^2 - 204\lambda \right)\lambda c_{\phi\square} - \frac{40}{3} g_1^2 c_{\phi D} \,,\nonumber\\[0.3cm]
    \beta_{c_{\phi D}}^{(2\ell)} &= -\frac{272}{3} g_1^2 c_{\phi\square} + \left(5g_1^2 - 144 \lambda\right)\lambda c_{\phi D} 
    \,,\nonumber\\[0.3cm]
    \beta_{c_{\phi}}^{(2\ell)} &= \left(144g_1^2 - 3444 \lambda\right)\lambda c_\phi + \left(-240g_1^2 + 7968 \lambda\right)\lambda^2c_{\phi\square}\,, \nonumber\\[0.3cm]
    &~~ - \left(6g_1^2 + 1992 \lambda\right)\lambda^2c_{\phi D}
    \,,\nonumber\\[0.3cm]
    \beta_{c_{\phi B}}^{(2\ell)} &= -4 g_1^2 \lambda c_{\phi\square} - 2 g_1^2 \lambda c_{\phi D} - 60\lambda^2 c_{\phi B}\,.
\end{align}
These RGEs match exactly the results presented in Ref.~\cite{Born:2024mgz}. Note that we use a different convention for $\lambda$. Note also that, in that reference, intermediate off-shell results are not provided. These can actually be of interest if the SM is extended with light particles, \textit{e.g.} ALPs or sterile neutrinos, not to repeat the computation from scratch.

Following the same approach, we compute the simple poles of the dimension-eight operators in Tab.~\ref{tab:smeftdim8} for vanishing $g_1$:
\begin{align}\label{eq:dim8divs}
    \tilde{K}_{\phi} &= 3\lambda^2\,,\nonumber\\[0.2cm]
    \tilde{\lambda} &= -84 \lambda^3\,,\nonumber\\[0.2cm]
    \tilde{r}_{\phi^2 D^6} &= -\frac{1}{8}\lambda(c_{\phi^4 D^4}^{(1)}+c_{\phi^4 D^4}^{(2)}+c_{\phi^4 D^4}^{(3)})\,,\nonumber\\[0.2cm]
    \tilde{c}_{\phi^4 D^4}^{(1)} &= \frac{1}{27}\lambda^2 (872 c_{\phi^4 D^4}^{(1)}+188c_{\phi^4 D^4}^{(2)}+160c_{\phi^4 D^4}^{(3)})\,,\nonumber\\[0.2cm]
    \tilde{c}_{\phi^4 D^4}^{(2)} &= \frac{1}{27}\lambda^2 (174 c_{\phi^4 D^4}^{(1)}+886c_{\phi^4 D^4}^{(2)}+160c_{\phi^4 D^4}^{(3)})\,,\nonumber\\[0.2cm]
    \tilde{c}_{\phi^4 D^4}^{(3)} &= \frac{1}{27} (554 c_{\phi^4 D^4}^{(1)}+410c_{\phi^4 D^4}^{(2)}+1396c_{\phi^4 D^4}^{(3)})\,,\nonumber\\[0.2cm]
    \tilde{r}_{\phi^4 D^4}^{(4)} &= \frac{1}{108}\bigg[\lambda (360 c_{\phi^6 D^2}^{(1)}-432 c_{\phi^6D^2}^{(2)})\nonumber\\
    &+\lambda^2 (893 c_{\phi^4 D^4}^{(1)}+547c_{\phi^4 D^4}^{(2)}+2088 c_{\phi^4 D^4}^{(3)})\bigg]\,,\nonumber\\[0.2cm]
    \tilde{r}_{\phi^4 D^4}^{(6)} &= \frac{1}{54}\bigg[\lambda (216 c_{\phi^6 D^2}^{(2)}-36 c_{\phi^6 D^2}^{(1)})\nonumber\\
    &-\lambda^2 (161 c_{\phi^4 D^4}^{(1)}+667 c_{\phi^4 D^4}^{(2)}+504 c_{\phi^4 D^4}^{(3)})\bigg]\,,\nonumber\\[0.2cm]
     \tilde{r}_{\phi^4 D^4}^{(8)} &= \frac{1}{36}\bigg[\lambda (6c_{\phi^6 D^2}^{(1)}-12 c_{\phi^6 D^2}^{(2)})\nonumber\\
    &-\lambda^2 (75c_{\phi^4 D^4}^{(1)}+65c_{\phi^4 D^4}^{(2)}+68c_{\phi^4 D^4}^{(3)})\bigg]\,,\nonumber\\[0.2cm]
     \tilde{r}_{\phi^4 D^4}^{(10)} &= \frac{1}{18}\bigg[\lambda (12 c_{\phi^6 D^2}^{(2)}-c_{\phi^6 D^2}^{(1)})\nonumber\\
    &-\lambda^2 (247c_{\phi^4 D^4}^{(1)}+277c_{\phi^4 D^4}^{(2)}+200c_{\phi^4 D^4}^{(3)})\bigg]\,,\nonumber\\[0.2cm]
     \tilde{r}_{\phi^4 D^4}^{(11)} &= \frac{1}{9}\bigg[\lambda (3c_{\phi^6 D^2}^{(1)}-30 c_{\phi^6 D^2}^{(2)})\nonumber\\
    &-\lambda^2 (29c_{\phi^4 D^4}^{(1)}+41c_{\phi^4 D^4}^{(2)}+34c_{\phi^4 D^4}^{(3)})\bigg]\,,\nonumber\\[0.2cm]
     \tilde{r}_{\phi^4 D^4}^{(12)} &= -\frac{1}{108}\bigg[\lambda (72 c_{\phi^6 D^2}^{(1)}+144 c_{\phi^6 D^2}^{(2)})\nonumber\\
    &+\lambda^2 (1171c_{\phi^4 D^4}^{(1)}+485c_{\phi^4 D^4}^{(2)}+1008 c_{\phi^4 D^4}^{(3)})\bigg]\,.\nonumber
\end{align}

We use now the field redefinitions in Refs.~\cite{Chala:2021cgt,Chala:2024llp}, that we cross-checked with \texttt{mosca}~\cite{LopezMiras:2025gar}:
\begin{align}
    \tilde{c}_{\phi^6 D^2}^{(1)} &\to \tilde{c}_{\phi^6 D^2}^{(1)} + 8\lambda^2 \tilde{r}_{\phi^2 D^6}+4 \lambda \tilde{r}_{\phi^4 D^4}^{(12)}-4\lambda \tilde{r}_{\phi^4 D^4}^{(4)}-2\lambda \tilde{r}_{\phi^4 D^4}^{(6)}\,,\nonumber\\
    \tilde{c}_{\phi^6 D^2}^{(2)} &\to \tilde{c}_{\phi^6 D^2}^{(2)} + 2 \lambda \tilde{r}_{\phi^4 D^4}^{(12)}-2\lambda \tilde{r}_{\phi^4 D^4}^{(6)}\,.
\end{align}
Physical four-derivative operators remain the same.
We work out the pure-two-loop anomalous-dimension matrix (ADM) $\gamma$, defined by $256\pi^4 \mu \dfrac{d c_i}{d\mu} = \gamma_{ij}c_j$. We obtain:
%
\begin{equation*}
   \resizebox{\columnwidth}{!}{$
   -\gamma = \left[
    \begin{array}{c|ccccc}
      & c_{\phi^4 D^4}^{(1)} & c_{\phi^4 D^4}^{(2)} & c_{\phi^4 D^4}^{(3)} & c_{\phi^6 D^2}^{(1)} & c_{\phi^6 D^2}^{(2)} \\[0.1cm] \hline\\[-0.2cm]
      c_{\phi^4 D^4}^{(1)} & {\color{gray}{\dfrac{2840}{27}\lambda^{2}}} & {\color{gray}{\dfrac{752}{27}\lambda^{2}}} & {\color{purplish}{\dfrac{640}{27}\lambda^{2}}} & {\color{blue}{0}} & {\color{blue}{0}} \\[0.3cm]
      c_{\phi^4 D^4}^{(2)} & {\color{gray}{\dfrac{232}{9}\lambda^{2}}} & {\color{gray}{\dfrac{2896}{27}\lambda^{2}}} & {\color{purplish}{\dfrac{640}{27}\lambda^{2}}} & {\color{blue}{0}} & {\color{blue}{0}} \\[0.3cm]
      c_{\phi^4 D^4}^{(3)} & \dfrac{2216}{27}\lambda^{2} & \dfrac{1640}{27}\lambda^{2} & \dfrac{4936}{27}\lambda^{2} & {\color{blue}{0}} & {\color{blue}{0}} \\ [0.3cm]
      c_{\phi^6 D^2}^{(1)} & \dfrac{38468}{27}\lambda^{3} & -\dfrac{16040}{27}\lambda^{3} & \dfrac{6608}{3}\lambda^{3} & \dfrac{4384}{3}\lambda^{2} & -\dfrac{352}{3}\lambda^{2} \\[0.3cm]
      c_{\phi^6 D^2}^{(2)} & {\color{greenish}{\dfrac{5584}{9}\lambda^{3}}} & {\color{greenish}{-\dfrac{5584}{9}\lambda^{3}}} & {\color{red}{0}} & {\color{red}{0}} & \dfrac{3856}{3}\lambda^{2}
    \end{array}\right]$}
\end{equation*}
%
A few comments regarding the colored values in the ADM are in order:

\begin{itemize}
\item
The vanishing entries in the first three lines ({\color{blue}{blue}}) are particularly interesting, because they hold only on-shell. (Namely, $\phi^4 D^4$ operators are renormalized by $\phi^6 D^2$ operators off-shell, as it is evident in Eqs.~\eqref{eq:dim8divs}.) This is, to the best of our knowledge, the best explicit evidence at dimension eight and two loops of the result in Ref.~\cite{Bern:2019wie} stating that operators with $n$ fields can not renormalize those with $m$ fields at $n-l$ loops or less. (In our case, $n=6$ and $m=4$.)

It is interesting to see that it can be also explained following the line of thought presented in Refs.~\cite{Chala:2023jyx,Chala:2023xjy}: In the deep IR, and based simply on dimensional grounds, $\phi^4 D^4$ WC's read schematically
\begin{align}
    c_{\phi^4 D^4}\sim \#\lambda c_{\phi^6 D^2}\log{\frac{\mu}{M_\Lambda}}+\cdots\,,
\end{align}
where \# is some numerical coefficients and the ellipses represent terms that are negligible in different UV completions of the SMEFT.

Now, $\phi^4 D^4$ operators are subject to positivity constraints~\cite{Remmen:2019cyz,Chala:2021wpj}:
\begin{align*}
c_{\phi^4 D^4}^{(2)} \geq 0 \,,  
\\ c_{\phi^4 D^4}^{(1)} + c_{\phi^4 D^4}^{(2)} \geq 0 \,,
\\ c_{\phi^4 D^4}^{(1)} + c_{\phi^4 D^4}^{(2)} + c_{\phi^4 D^4}^{(3)} \geq 0 \,.
\end{align*}
So for these equations to hold at $\mu\ll M_\Lambda$, it must be the case that $\# \lambda c_{\phi^6 D^2}\leq 0$. But since neither $\lambda$ nor $c_{\phi^6 D^2}$ are restricted by positivity, it should necessarily be $\# = 0$.

\item
It is also worth highlighting the two zeroes in the last row ({\color{red}{red}}). They follow from custodial symmetry. The operators $\mathcal{O}_{\phi^4 D^4}^{(1)}$ and $\mathcal{O}_{\phi^4 D^4}^{(2)}$ break custodial symmetry independently, while $\mathcal{O}_{\phi^4 D^4}^{(3)}$ and $\mathcal{O}_{\phi^6 D^2}^{(1)}$ do not.~\footnote{This is trivial to show. Simply consider the bidoublet defined by $\Phi = (i\sigma_2\phi,\phi)$, transforming as $\Phi\to L^\dagger\Phi R$ under the custodial symmetry group $ SU(2)_L\times SU(2)_R$. Then, $$D_\mu\phi^\dagger D_\nu\phi = \frac12 \text{tr}[D_\mu\Phi^\dagger D_\nu\Phi]-\frac12 \text{tr}[D_\mu\Phi^\dagger D_\nu\Phi\sigma_3]\,.$$ The first part on the RHS is invariant, while the second is not but vanish for $\mu=\nu$.} So, the latter cannot renormalize the also custodial-breaking operator $\mathcal{O}_{\phi^6 D^2}^{(2)}$. 

\item
Moreover, the particular combination $\mathcal{O}_{\phi^4 D^4}^{(1)}+\mathcal{O}_{\phi^4 D^4}^{(2)}$ preserves custodial too, from where we understand the relation between the first two entries in the last row of the ADM ({\color{greenish}{green}}). Should the WC's of these two operators be set equal, the custodial-breaking operator $\mathcal{O}_{\phi^6 D^2}^{(2)}$ will not get renormalized.

\item
For this same reason, the first two entries in the third column ({\color{purplish}{purple}}) must be equal. The custodial-symmetric $\mathcal{O}_{\phi^4 D^4}^{(3)}$ operator must renormalize $\mathcal{O}_{\phi^4 D^4}^{(1)}$ ans $\mathcal{O}_{\phi^4 D^4}^{(2)}$ in a custodial-preserving way. 

\item 
Another crosscheck can be done regarding the top left $2\times2$ sub-matrix ({\color{gray}{gray}}). If we add up the entries in the first row and the entries in the second row, the result should be the same, namely,
$$\frac{2840}{27}\lambda^2 + \frac{752}{27}\lambda^2
=
\frac{232}{9}\lambda^2 + \frac{2896}{27}\lambda^2 \,.
$$
This way, if the operator $\mathcal{O}_{\phi^4 D^4}^{(1)}+\mathcal{O}_{\phi^4 D^4}^{(2)}$ is turned on, the RG flow preserves this specific combination.

\end{itemize}

\subsection{CFT spectrum in the Abelian Higgs model}
\label{subsec:abelian higgs}
Let $\Phi=(\Phi_1,\dots,\Phi_n)$ be $n$ complex charged scalars. We are interested in the following Lagrangian:
\begin{align}
    \mathcal{L} = D_\mu \Phi^\dagger D^\mu\Phi - m^2  \Phi^\dagger\Phi - \lambda (\Phi^\dagger\Phi)^2\,,
\end{align}
where
\begin{equation}
    D_\mu\Phi = \partial_\mu\Phi + igA_\mu\Phi\,.
\end{equation}

This model has a charged fixed-point in $d=3$ dimensions at large $n$, visible within $\epsilon$-expansion around $d=4-2\epsilon$.~\footnote{We adopt in this paper the $4-2\epsilon$ convention instead of the usual $4-\epsilon$ expansion for the study of the fixed-points. See Ref.~\cite{Ihrig:2019kfv} for more details.} In what follows, we compute this as well as scalings produced by irrelevant interactions shown in Tab.~\ref{tab:dim6ops}
\begin{table}[t]
 \begin{tabular}{|p{0.3\columnwidth} m{0.5\columnwidth}|}
 \hline
 \rowcolor{gray!20} \multicolumn{2}{|c|}{$\boldsymbol{\Phi^2 D^4}$} \\
 \hline
 \hspace{0.1\columnwidth} \textcolor{gray}{$\mathcal{O}_{\Phi^2 D^4}$} & \textcolor{gray}{$D^2\Phi^\dagger D^2\varphi$} \\
 \hline
 \rowcolor{gray!20} \multicolumn{2}{|c|}{$\boldsymbol{\Phi^4 D^2}$} \\
 \hline
 \hspace{0.1\columnwidth} $\mathcal{O}_{\Phi^4 D^2}^{(1)}$ & $(\Phi^\dagger \Phi) (D_\mu\Phi^\dagger D^\mu \Phi)$ \\
 \hspace{0.1\columnwidth} \textcolor{gray}{$\mathcal{O}_{\Phi^4 D^2}^{(2)}$} & \textcolor{gray}{$(\Phi^\dagger\Phi) (D^2\Phi^\dagger\Phi+\Phi^\dagger D^2\Phi)$} \\
 \hspace{0.1\columnwidth} \textcolor{gray}{$\mathcal{O}_{\Phi^4 D^2}^{(3)}$} & \textcolor{gray}{$\ii(\Phi^\dagger \Phi)(D^2\Phi^\dagger\Phi+\Phi^\dagger D^2\Phi)$} \\
 \hline
 \rowcolor{gray!20} \multicolumn{2}{|c|}{$\boldsymbol{\Phi^6}$} \\
 \hline
 \hspace{0.1\columnwidth} $\mathcal{O}_\Phi^6$ & $(\Phi^\dagger\Phi)^3$\\
 \hline
 \rowcolor{gray!20} \multicolumn{2}{|c|}{$\boldsymbol{X^2 D^2}$} \\
 \hline
 \hspace{0.1\columnwidth} \textcolor{gray}{$\mathcal{O}_{2A}$} & \textcolor{gray}{$-\frac{1}{2}\partial_\mu A^{\mu\nu} \partial_\rho A^{\rho\nu}$}\\
 \hline
 \rowcolor{gray!20} \multicolumn{2}{|c|}{$\boldsymbol{\Phi^2 X D^2}$} \\
 \hline
 \hspace{0.1\columnwidth} \textcolor{gray}{$\mathcal{O}_{AD\Phi}$} & \textcolor{gray}{$D_\nu A^{\mu\nu} (\Phi^\dagger \ii \overleftrightarrow{D}_\mu\Phi)$}\\
 \hline
 \rowcolor{gray!20} \multicolumn{2}{|c|}{$\boldsymbol{X^2\Phi^2}$} \\
 \hline
 \hspace{0.1\columnwidth} $\mathcal{O}_{\Phi A}$ & $(\Phi^\dagger\Phi) A_{\mu\nu} A^{\mu\nu}$\\
 \hline
 \end{tabular}
 \caption{\it Dimension-six operators in the Abelian Higgs model.}\label{tab:dim6ops}
\end{table}
For the fix point, we take the beta function for $g^2$ and $\lambda$ up to two loops from the automated tool \texttt{RGBeta}~\cite{Thomsen:2021ncy}, and cross-check using our approach:
\begin{align}
    \beta_{g^2} &= -2 \epsilon g^2 + \frac{n}{24\pi^2} g^4 + \frac{n}{32\pi^4} g^6\,,\\
    \beta_{\lambda} &= -2\epsilon\lambda + 
    \frac{1}{16\pi^2} \left[
    6g^4 - 12 g^2\lambda + 4 (4+n)\lambda^2
    \right]
    \nonumber \\ &
    + \frac{1}{256\pi^4} \bigg[
    -\frac{4(45+7n)g^6}{3} + \frac{174+142n}{3} g^4\lambda
    \nonumber \\ &
    + (80+32n) g^2 \lambda^2 - 24 (7+3n) \lambda^3
    \bigg]
    \,.
\end{align}
We can safely ignore the running of $m^2$, in the limit of massless $\Phi$ relevant at the fixed point.

These two beta functions vanish, in particular~\footnote{There is also a trivial Gaussian fixed point as well as the Wilson-Fisher point of pure scalar theory. All of them are neutral (namely $g_*=0$), while we are interested in charged ones.}, at:
\begin{align}
    g_*^2 &= \frac{48\pi^2}{n} \left(
        \epsilon - \frac{36}{n}\epsilon^2
    \right)\,,\nonumber\\
    \lambda_* &= -\frac{4 \pi^2}{n^2(4+n)^3 b^2} \left[
    p(n,b) \epsilon + q(n,b) \epsilon^2
    \right]\,,
    \label{eq:fixed points formulas}
\end{align}
where $b=\sqrt{n^2-180n-540}$ and $p,q$ are two sixth-order polynomials in $n$ and $b$. 

The theory around this fixed point is conformal. The dimension-six singlet spectrum of the corresponding CFT is provided by the eigenvalues of the ADM involving the operators in Tab.~\ref{tab:dim6ops}.

This fixed point provides only real values for $\lambda$ when $b^2>0$, which occurs for $n\gtrsim 183$. Since $\lambda_*$ presents a divergent behavior close to this region of the parameter space, we will be interested in the large-$n$ limit, where $\lambda_*,g_*$ are small enough not to spoil perturbation theory. In this approximation, and setting $\epsilon=1/2$ so that we work in $d=3$ dimensions:
\begin{align}\label{eq:fixedpoint}
   g_*^2 &= \frac{24\pi^2}{n} - \frac{432\pi^2}{n^2}\,,\nonumber\\
    \lambda_* &= -\frac{288\pi^2}{n^2} - \frac{25\,596\pi^2}{n^3} - \frac{3\,367\,980\pi^2}{n^4} + \mathcal{O}\left(\frac{1}{n^5}\right)\,.
\end{align}
Notice that $g_*^2 > |\lambda_*|$ for any legit value of $n$. However, in the large-$n$ limit, the effective quartic coupling is $n \lambda$,
and so $|n \lambda_*|\gg g_*^2$, implying that $\mathcal{O}(g^4)$ contributions in the ADM can be safely neglected. (Moreover, it happens that $\lambda$ terms in the ADM are numerically enhanced with respect to $g$ ones.)

In such approximation, we obtain:
\begin{align}
    \tilde{r}_{\Phi^2 D^4} &=  \frac{1}{256\pi^4}\left(
    \frac{5}{12}g^2 + \frac{1+n}{6} \lambda
    \right) c_{\Phi^4 D^2}^{(1)}\,,\nonumber\\[0.1cm]
    \tilde{c}_{\Phi^4 D^2}^{(1)} &= \frac{1}{16\pi^2}\left[
    -3 g^2 - 4\left(1+n\right) \lambda
    \right] c_{\Phi^4 D^2}^{(1)}  
    \nonumber\\ &\quad + 
    \frac{1}{256\pi^4}\left[
    -\frac{55+15n}{3}g^2\lambda - 17\left(1+n\right) \lambda^2 
    \right] c_{\Phi^4 D^2}^{(1)}\,,\nonumber\\[0.1cm]
    \tilde{r}_{\Phi^4 D^2}^{(2)} &= \frac{1}{16\pi^2}
    \left(1-n\right) \lambda c_{\Phi^4 D^2}^{(1)}
    + \frac{1}{256\pi^4}\bigg[
    -\left(6+3n\right) c_{\Phi^6} 
    \nonumber\\&\quad
    -\left(\frac{125-6n}{12} g^2 \lambda +
    \frac{1-n}{2} \lambda^2
    \right)c_{\Phi^4 D^2}^{(1)}
    \bigg]\,,\nonumber\\[0.1cm]
    %
    %
    \tilde{c}_{\Phi^6} &= \frac{1}{16\pi^2}\left\{
    \left[9g^2 - 6(7+n)\lambda\right] c_{\Phi^6}
    - 4(4+n)\lambda^2 c_{\Phi^4 D^2}^{(1)}
    \right\} 
    \nonumber\\&\quad
    + \frac{1}{256\pi^4}\bigg\{
    \left[24(4+n)g^2 \lambda -3(197+53n)\lambda^2\right] c_{\Phi^6}
    \nonumber\\&\quad
     + \left[
     -16(1-n) g^2\lambda^2 - 12 (29+5n)
     \right] c_{\Phi^4 D^2}^{(1)}
    \bigg\}\,,\nonumber\\[0.1cm]
    %
    %
    \tilde{r}_{AD\Phi} &= \frac{1}{96\pi^2}g c_{\Phi^4 D^2}^{(1)} + \frac{5}{4608\pi^4}\left(
    1+n
    \right)g\lambda c_{\Phi^4 D^2}^{(1)}\,,\nonumber\\[0.1cm]
    \tilde{c}_{\Phi A} &= -\frac{1}{8\pi^2}\left(
    1+n
    \right)\lambda c_{\Phi A} 
    \nonumber\\&\quad 
    - \frac{1}{256\pi^4}\left[
    \frac{2}{3}(1+n) g^2 \lambda c_{\Phi^4 D^2}^{(1)}
    + 5(1+n) \lambda^2 c_{\Phi A}
    \right]\,,
\end{align}
where all other vanish.

Using again \texttt{mosca}~\cite{LopezMiras:2025gar} for deriving the field redefinitions (see also Ref.~\cite{Bernardo:2025vkz}),
\begin{align*}
    \tilde{c}_{\Phi^4 D^2}^{(1)}&\to \tilde{c}_{\Phi^4 D^2}^{(1)}-3g^2 \tilde{r}_{2A}+6g\tilde{r}_{AD\Phi}\,,\nonumber\\
\tilde{c}_{\Phi^6}&\to\tilde{c}_{\Phi^6}+2g_1^2\lambda\tilde{r}_{2A}-4g\lambda \tilde{r}_{AD\Phi}+4\lambda^2\tilde{r}_{\Phi^2D^4}^{(1)}-4\lambda\tilde{r}_{\Phi^4 D^2}^{(2)}\,,\nonumber\\
\tilde{c}_{\Phi A}&\to \tilde{c}_{\Phi A}\,,
\end{align*}
we obtain the following ADM at the fixed point:
\small
\begin{equation*}
   \gamma = \left[
    \begin{array}{c|ccc}
      & c_{\Phi^4 D^2}^{(1)} & c_{\Phi^6} & c_{\Phi A}  \\[0.1cm] \hline\\[-0.2cm]
      c_{\Phi^4 D^2}^{(1)} & 
      - \frac{273}{2n} - \frac{\color{blue}{12\,447}}{n^2}
       & 0  & 0   \\[0.3cm]
      c_{\Phi^6} & \frac{576\pi^2}{n^3} & -\frac{243}{n} -\frac{\color{blue}{22\,815}}{n^2} & 0 \\[0.3cm]
      c_{\Phi A} & 0 & 0 & -\frac{72}{n}-\frac{6\,471}{n^2}
    \end{array}\right]\,.
    %
\end{equation*}
\normalsize
We only show the first terms in an asymptotic $1/n$ expansion. Also, we represent in \textcolor{blue}{blue} those numbers that change with respect to considering only one-loop beta functions.

\begin{figure}[t]
  \includegraphics[width=\linewidth]{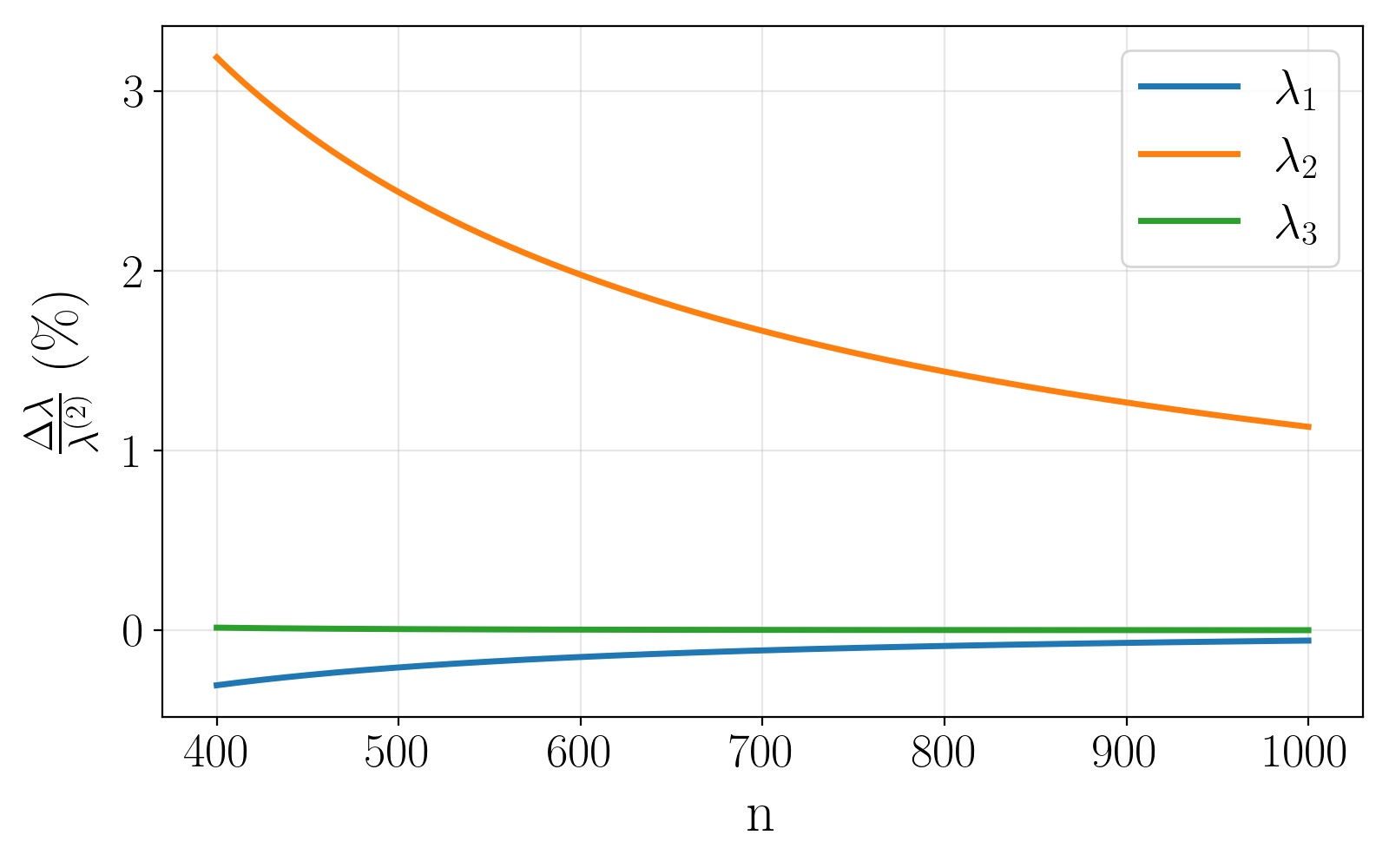}
  \caption{\it Relative difference (in percent) between the the eigenvalues of the dimension-six singlet ADM in the large-flavor Abelian Higgs model up to one- and two-loop order.}\label{fig:cft}
\end{figure}

The eigenvalues of $\gamma$ are, up to order $\mathcal{O}(1/n^4)$,
\begin{align*}
    \lambda_1 &= -\frac{267}{2n}-\frac{12\,681}{n^2}-\frac{6\,740\,145}{n^3}\,,
    \\
    \lambda_2 &= -\frac{273}{n}-\frac{12\,447}{n^2}-\frac{6\,688\,395}{n^3}\,,
    \\
    \lambda_3 &= -\frac{72}{n}-\frac{6\,471}{n^2}-\frac{854\,874}{n^3}\,.
\end{align*}
We show in Fig.~\ref{fig:cft} the relative difference $\Delta\lambda/\lambda^{(2)}$, where $\Delta\lambda = \lambda^{(1)}-\lambda^{(2)}$ and $\lambda^{(1,2)}$ are the ADM eigenvalues computed from up-to-one-loop and up-to-two-loop beta functions, respectively. Including two-loop effects can modify the results by as much as 3\%.

\section{Conclusions}
\label{sec:conclusions}
We have proposed a novel way of extracting two-loop divergences without sub-divergence contamination, different from previous approaches based on propagator deformations, infrared re-arrangement of $R^*$.

Our approach consists in viewing a 4D EFT as the IR limit of the 5D counterpart compactified on a circle. The 4D UV divergences ensue as the IR divergences in the matching in dimensional reduction. Doing so, we have been able to cross-check recent results in dimension-six SMEFT running at two loops, as well as to renormalize the scalar sector of the dimension-eight SMEFT to two loops and compute the anomalous dimensions of the large-flavor Abelian Higgs model at the charged fixed point.

Our approach thus connects different areas of high-energy physics, including EFT, extra dimensions and sum-integral research in thermal field theory. Interestingly, because two-loop sum-integrals factorize into one-loop sum-integrals (this is in contrast with regular two-loop integrals), our method suggests that all data relevant for two-loop running is encoded in simple loops.

Our results, so far, restrict to bosonic EFT, because the process of compactification removes zeroth modes of fermionic fields. It would therefore be interesting to explore other compactification schemes, \textit{e.g.} on orbifolds.

\vspace{0.2cm}

\textbf{Note added:} During the completion of this work, Ref.~\cite{Henriksson:2025vyi} appeared on the arXiv, in which SMEFT two-loop dimension-eight RGEs are also computed. However, they use a rather different basis of operators, making the comparison highly non-trivial.

\appendix

\section{Explicit example to higher-order in gauge coupling}\label{app:example}
For our method to work, it is crucial that the IR limit of the 5D theory has the same content ---both in fields and in operators--- as the theory that we ultimately want to renormalize. Mathematically speaking, let $$\mathcal{L}(c_i) = \sum_i c_i \mathcal{O}_i$$ be the Lagrangian of the theory of which we want to compute the counterterms, expressed as a function of the WC's. Also define analogously $\mathcal{L}'(c'_i)$ as the Lagrangian of the 4D EFT after the dimensional reduction procedure. Then, as long as $$\mathcal{L}'(c_i) = \mathcal{L}(c_i)\,,$$ and provided we know the relations $c'_i=c'_i(c_j)$ driven by the one-loop finite matching equations, we can extract the divergences of $\mathcal{L}(c_i)$ from those of $\mathcal{L}'(c'_i)$.

The simple fact of considering a higher-dimensional version of the theory \textit{almost} does the trick. 
The issue is that, in the presence of gauge interactions, the 5D version reduces to the 4D with a new scalar coming from the fifth component of the gauge vector. Said differently, take any (bosonic) 4D theory with Lagrangian $\mathcal{L}$, promote it to 5D and compute the 4D effective Lagrangian $\mathcal{L}'$ upon dimensional reduction; then, the 4D EFT is identical in content of fields and operators to the original theory, but with an extra real scalar for every gauge symmetry. Namely,
$$
\mathcal{L}'(c_i) = \mathcal{L}(c_i) + \Delta\mathcal{L}(c_i)\,,
$$
where $\Delta\mathcal{L}(c_i)$ only contains operators involving the new scalars. Therefore, our method does not truly compute the divergences of the original 4D theory, but rather those of the scalar-extended one.

We can explore this a bit further. Let us call $T$ to the theory from which we want to extract two-loop divergences. We represent its Lagrangian as $\mathcal{L}_T(c_i)$. For the sake of simplicity, we will assume that $T$ involves a single gauge symmetry $G$, with gauge vector $A_\mu$ and gauge coupling $g$. Now, let $T'$ be the 4D EFT theory corresponding to 5D analog of $T$ at the one-loop level. $T'$ has the same content in fields as $T$ as well as an extra real scalar field $A_5$ transforming in the adjoint representation of $G$. This new field comes, at tree-level, either from operators of $T$ with covariant derivatives ---in which case $A_5$ needs to come in pairs with explicit factors of $g^2$--- or from operators involving $A^{\mu\nu}$. Then, in general, the two theories $T$ and $T'$ differ by 
\begin{equation}
    \mathcal{L}_{T'}(c_i) = \mathcal{L}_{T}(c'_i) + \mathcal{L}_{A_5}(c_i) \,
\end{equation}
where the Lagrangian containing $A_5$ is of order $\mathcal{O}\left(g^2,c_{A}\right)$ and $c_A$ denotes any WC of an operator in $\mathcal{L}_T$ containing $A_{\mu\nu}$. This still holds at the loop level.

Our procedure is then engineered to compute the divergences of $T'$, not those of $T$. Namely, it is also taking into account how operators from $\mathcal{L}_{A_5}$ mix into those of $\mathcal{L}_{T}$. However, for this to happen, there needs to be at least two operators from $\mathcal{L}_{A_5}$ in the mixing, so that the $A_5$ field remains inside the loops. \footnote{With the exception of tadpoles, where $A_5$ can be in a loop and the diagrams only contribute with a single operator from $\mathcal{L}_{A_5}$. However, massless tadpoles do not generate divergences.} Therefore, this contribution is always order $g^4, g^2 c_A$ or $c_A^2$. More generally, if we include $g^2\in \{c_A\}$, we can correctly derive from our approach contributions up to order $\mathcal{O}(c_A^2)$.

It turns out that there is a straightforward way to correct contributions coming with higher powers of $c_A$. One can compute the divergences of the theory $T'$ (only those contributing to operators of $T$ though), keep those coming from insertions of $A_5$, and subtract them from the previous divergences. Of course, in order for our method to be self-contained, the same formalism could be used to compute these divergences. 
So, we would look for the divergences of $T'$ by promoting $\mathcal{L}_T'$ to 5D and performing the dimensional reduction. From our previous reasoning, we know that the obtained divergences would not be those of $T'$, but rather those of some theory $T''$ which couples $T'$ to a new scalar $A_5'$. Fortunately, we could ignore this up to order $\mathcal{O}(c_A^2)$. Since $A_5$ operators came at least with order $c_A$, we could trust these divergences up to order $\mathcal{O}(c_A^3)$. After subtracting them from the original results, the right contributions from order $c_A^2$ are recovered.

To illustrate this whole point, we will show the process of retrieving the divergence of $c_{\phi\square}$ in \ref{subsec:rges smeft} coming from $g_1^2 c_{\phi B}$, for which we will need to take into account the subtraction of $B_5$ contributions. In order to focus in this particular contribution, consider the simplified Lagrangian
\begin{equation*}
    \mathcal{L}_T = -\frac{1}{4}B_{\mu\nu}B^{\mu\nu} + |D_\mu\phi|^2 - \lambda |\phi|^4 + c_{\phi B} B_{\mu\nu}B^{\mu\nu} |\phi|^2 \,,
\end{equation*}
where $c_{\phi B}$ has mass dimension $[c_{\phi B}]=-2$
and $g_1$ is the gauge coupling.

We now write $\mathcal{L}_T$ in 5D, which just requires the following exchanges: $g_1 \to g_1/\sqrt{\Lambda}$, $\lambda \to \lambda/\Lambda$ and $c_{\phi B} \to c_{\phi B}/\Lambda$, where $1/\Lambda$ is the size of the fifth dimension.

After performing the two-loop matching to the 4D EFT (with no renormalization of the UV theory), the Lagrangian reads
\begin{align}
    \mathcal{L}_{T'} &= 
    -\frac{K_B}{4}B_{\mu\nu}B^{\mu\nu} + K_H |D_\mu\phi|^2 - C_\lambda |\phi|^4 + 
    \nonumber \\ & \quad C_{\phi B} B_{\mu\nu}B^{\mu\nu} |\phi|^2
    + C_{\phi\square} |\phi|^2\square|\phi|^2 + \ldots \,
    \label{eq:lag 4d eft example}
\end{align}
where the ellipses include other operators that will play a role afterwards. Furthermore, we can split every EFT coefficient $C_i$ into a finite part ---that, with a little abuse of notation, we will call $C_i$--- and $1/\epsilon^n$ divergent parts $\delta_n C_i$. The relevant part for our purposes reads:
\begin{align}
    C_{\phi \square} &= \frac{\hbar}{16\pi^2} \left(
    \frac{1}{18} \frac{g_1^2 \lambda}{\Lambda^2} - \frac{5}{36} \frac{\lambda^2}{\Lambda^2} - g_1^2 c_{\phi B} 
    \right) + \mathcal{O}(\hbar^2) \,, \nonumber \\
    \delta_1 C_{\phi \square} &= \frac{\hbar}{8\pi^2\epsilon} g_1^2 c_{\phi B} + \frac{\hbar^2}{256\pi^4\epsilon} \left(
    \frac{77}{30} \frac{\lambda^3}{\Lambda^2} + \frac{3}{2} g_1^2 c_{\phi B} 
    \right) \,.
    \label{eq:matching results for chbox}
\end{align}

We have kept the factors of $\hbar$ to make explicit the one- and two-loop nature of the different terms. Naively, one would read $\delta_1 C_{\phi \square}$ as the divergence of $c_{\phi \square}$. The main reason why this is not the case is because of the $\lambda^3$ factor renormalizing a dimension-six operator which, on dimensional grounds, is not possible in 4D but it is allowed in 5D. (Namely, the presence of $\Lambda$ already indicates that this is inherently a 5D effect). Hence, this is an artifact arising from the fact that we did not introduce the two-loop counterterms of the 5D theory. Luckily, they are not needed whatsoever because we can trivially discard these terms on the basis of their energy dimensions.

Moreover, in $\delta_1C_i$ we find two-loop divergences of the tree-level Lagrangian (which is what we are looking for) together with one-loop divergences of the one-loop-generated finite Lagrangian. That is to say, the finite value of $C_{\phi\square}$ yields, when inserted in one-loop diagrams, a divergence of order $\mathcal{O}(\hbar^2/\epsilon)$, which is already taken into account in $\delta_1 C_{\phi\square}$ and we have to subtract. It turns out that the only non-zero one-loop finite part in $\mathcal{L}_{T'}$ consistent with the 4D mass dimensions is the $g_1^2 c_{\phi B}$ term in $C_{\phi\square}$ in Eq.~\eqref{eq:matching results for chbox}. At the one-loop level, the self-renormalization of $C_{\phi\square}$ is
\begin{equation*}
    \widetilde C_{\phi\square} \supset \frac{\hbar}{16\pi^2\epsilon} \left(
    \frac{1}{4}g_1^2 + 14\lambda 
    \right)
    C_{\phi \square} \,.
\end{equation*}
After replacing the finite one-loop part of $C_{\phi\square}$, we find the term
\begin{equation}
   (\Delta \delta_1 C_{\phi\square})_1 = -\frac{7\hbar^2}{128\epsilon} g_1^2 \lambda c_{\phi B} + \mathcal{O}(g_1^4)\,,
   \label{eq:two loop divergence extra contribution 1}
\end{equation}
which we have to subtract from $\delta_1 C_{\phi B}$.~\footnote{Notice that this process is equivalent to realizing that $\delta_1 C_{\phi\square}$ is the divergence of $C_{\phi\square}$ rather than that of $c_{\phi\square}$. Then the one-loop change of variables $c_i=c_i(C_j)$ is required.}

The subtraction of \eqref{eq:two loop divergence extra contribution 1} to \eqref{eq:matching results for chbox} does not give the right result yet. One may recall that in the ellipses in Eq.~\eqref{eq:lag 4d eft example} there are also operators with a new singlet scalar $B_5$. At tree-level we find:
\begin{multline*}
    \mathcal{L}_{A_5} \supset C_{\phi^2 B_5^2} |\phi|^2 B_5^2 + C_{D^2 \phi^2 B_5^2}^{(1)} |\phi|^2 B_5 \square B_5
    \\ 
    + C_{D^2 \phi^2 B_5^2}^{(2)} (B_5^2\phi^\dagger D^2\phi + \mathrm{h.c.}) + C_{D^2 \phi^2 B_5^2}^{(3)} |D_\mu \phi|^2 B_5^2 \,,
\end{multline*}
with $C_{\phi^2 B_5^2} = \frac{1}{4}g_1^2$ as well as $C_{D^2 \phi^2 B_5^2}^{(1)}=-2C_{D^2 \phi^2 B_5^2}^{(2)}=-C_{D^2 \phi^2 B_5^2}^{(3)} = -2c_{\phi B}$. These also mix into $C_{\phi\square}$ via two-loop diagrams. In particular,
\begin{align}
    \widetilde C_{\phi\square} &\supset -\frac{5\hbar^2}{512\pi^4\epsilon} \lambda C_{\phi^2 B_5^2} 
    \nonumber\\ & \qquad
    \quad\times\left(
    C_{D^2\phi^2 B_5^2}^{(1)}
    +2C_{D^2\phi^2 B_5^2}^{(2)}+C_{D^2\phi^2 B_5^2}^{(3)}   
    \right)
    \nonumber\\ &
    = -\frac{5\hbar^2}{1024\pi^2\epsilon} g_1^2 \lambda c_{\phi B} \equiv (\Delta \delta_1 C_{\phi\square})_2\,.
    \label{eq:two loop divergence extra contribution 2}
\end{align}
A similar procedure applies to the one-loop generated $B_5$ operators which could mix into $C_{\phi\square}$ at the one-loop level. These have at most two $B_5$ legs. The only relevant contribution is
\begin{equation}
    \widetilde C_{\phi\square} \supset -\frac{\hbar}{8\pi^2\epsilon} C_{\phi^2 B_5^2}  C_{D^2\phi^2 B_5^2}^{(2)} \,.
\end{equation}
After the replacement for the one-loop finite parts,
\begin{align*}
    C_{\phi^2 B_5^2} &= \frac{1}{4}g_1^2 + \frac{3\hbar}{16\pi^2}\lambda g_1^2 \,,
    \\
    C_{D^2 \phi^2 B_5^2}^{(2)} &= c_{\phi B} + \frac{\hbar}{16\pi^2}g_1^2 c_{\phi B} \,,
\end{align*}
yields
\begin{equation}
    (\Delta \delta_1 C_{\phi\square})_3 = -\frac{\hbar}{32\pi^2\epsilon}g_1^2 c_{\phi B} - \frac{3\hbar^2}{128\pi^4\epsilon} g_1^2\lambda c_{\phi B} + \mathcal{O}(g_1^4) \,.
    \label{eq:two loop divergence extra contribution 3}
\end{equation}
Here we find also a modification of the one-loop divergence in Eq.~\eqref{eq:matching results for chbox}. 

By subtracting the contributions of Eqs.~\eqref{eq:two loop divergence extra contribution 1}, \eqref{eq:two loop divergence extra contribution 2} and \eqref{eq:two loop divergence extra contribution 3} from \eqref{eq:matching results for chbox} we finally get the correct two-loop simple pole factor of $c_{\phi\square}$, namely
\begin{equation}
    \tilde c_{\phi\square} = \frac{91\hbar^2}{1024\pi^4\epsilon}g_1^2\lambda c_{\phi B}\,.
\end{equation}
In accordance with the field redefinitions in Eq.~\eqref{eq:field redefinitions smeft 6}, we repeat the process to compute the relevant divergences of $r_{\phi D}',r_{2B}$ and $r_{BD\phi}$:
\begin{equation*}
    \tilde{r}_{\phi D}'=-\frac{7\hbar^2}{512} g_1^2 \lambda c_{\phi B} \,, \qquad
    \tilde{r}_{BD\phi} = -\frac{3\hbar^2}{128\pi^4\epsilon} g_1 \lambda c_{\phi B} \,
\end{equation*}
and $\tilde{r}_{2B} = 0$. After combining all together, the divergence of $c_{\phi \square}$ in the on-shell basis reads
\begin{equation}
    \tilde c_{\phi\square} = \frac{9\hbar^2}{128\pi^4\epsilon}g_1^2\lambda c_{\phi B}\,,
\end{equation}
in agreement with the beta function in Ref.~\cite{Born:2024mgz}.

\section*{Acknowledgments} 
We thank Javier Fuentes-Martín, Luis Gil and Jose Santiago for useful discussions. We acknowledge support from the MCIN/AEI (10.13039/501100011033) and ERDF (grants PID2022-139466NB-C21/C22), from the Junta de Andaluc\'ia grants FQM 101 and P21-00199 as well as from Consejer\'ia de Universidad, Investigaci\'on e Innovaci\'on, Gobierno de Espa\~na and Uni\'on Europea - NextGenerationEU/PRTR (grants AST22 6.5 and EUR2025.164833). JLM is further supported by the FPU program under grant number FPU23/02028.


\bibliographystyle{apsrev4-2}
\bibliography{refs}

\end{document}